\documentclass[useAMS,usenatbib,usegraphicx]{mn2e}

\title[]{Infrared studies of the Be star X Per}
\author[Mathew et al.]{Blesson Mathew\thanks{E-mail:blesson@prl.res.in},
  D. P. K. Banerjee, Sachindra Naik and N. M. Ashok\\
Astronomy and Astrophysics Division, Physical Research Laboratory,
  Navrangapura, Ahmedabad - 380 009, Gujarat, India}

\begin{document}


\maketitle

\label{firstpage}

\begin{abstract}
Photometric and spectroscopic results are presented for the
Be star X Per/HD 24534 from near-infrared monitoring in 2010 -- 2011.
The star is one of a sample of selected Be/X-ray binaries being monitored by
us in the near-IR to study correlations between their X ray and near-IR
behaviour. Comparison of the star's present near-IR magnitudes with
earlier records shows the star to be currently in a prominently bright
state with mean $J, H, K $ magnitudes of 5.49, 5.33 and 5.06 respectively.
The $JHK$ spectra are dominated by emission lines of He{\sc i}
and Paschen and Brackett lines of H{\sc i}.
Lines of O{\sc i} 1.1287 and 1.3165 $\mu$m are also present and their relative
strength indicates, since O{\sc i} 1.1287 is stronger among the two lines,
that Lyman $\beta$ fluorescence plays an important role in their excitation.
Recombination analysis of the H{\sc i} lines is done which shows
that the Paschen and Brackett line strengths deviate considerably from case B predictions.
These deviations are attributed to the lines being optically thick and
this supposition is verified by calculating the line center optical depths
predicted by recombination theory. Similar calculations indicate that the
Pfund and Humphrey series lines should also be expected to be optically
thick which is found to be consistent with observations
reported in other studies. The spectral energy distribution of the star
is constructed and shown to have an infrared excess.
Based on the magnitude of the IR excess, which is modeled using a free-free
contribution from the disc, the electron density in the disc is estimated
and shown to be within the range of values expected in Be star discs.

\end{abstract}


\section{Introduction}
The Be star X Per/HD 24534 is the optical/IR counterpart of the X-ray source
4U0352+30 and belongs to the class of Be/X-ray binaries.
The orbital parameters of the system have been estimated by \citet{DelgadoMarti01} as follows:
a period of 250 days, an eccentricity of 0.11 and an inclination angle  between 26 and
33 degrees. \citet{White76} found evidence of 13.9 min modulations of the
X-ray flux using the data taken with Copernicus and Ariel 5 satellites.
\citet{DelgadoMarti01} found similar modulation of 837 s in RXTE data,
which possibly corresponds to the spin period of the neutron star.
The Be star was classified to be of O9.5 {\sc iii}e type with a
rotation velocity ($v~sini$) of 200 km s$^{-1}$ and lying at a distance of
1300 $\pm$ 400 pc \citep{Slettebak82,Norton91}.
\citet{Lyubimkov97} re-estimated the spectral type, $v~sini$
and distance using the data taken during a low-luminosity disc-less phase (1989-91) to be
B0Ve, 215 $\pm$ 10 km s$^{-1}$ and 700 $\pm$ 300 pc, respectively. 
From optical and infrared photometric data spanning a decade (1987-95),
\citet{Roche97} estimated the spectral type and distance as B0V and
900 $\pm$ 300 pc, respectively, during the disc-less phase.

From optical spectroscopy and infrared photometry during the period 1988-90,
\citet{Norton91} identified the loss of the circumstellar disc in X Per.
This was based on the change of the H$\alpha$ profile from emission to absorption,
an associated decrease in the infrared flux and the flattening of the infrared spectrum.
\citet{Fabregat92} used this dataset to study the astrophysical parameters of
X Per since the loss of disc revealed the normal B-type star. They estimated the
spectral type of the star to be O9.5 {\sc iii}e and set a lower age limit of 6 Myr for X Per
system. From high resolution optical spectroscopy and $V$ band photometry,
\citet{Clark01} identified an episode of complete disc loss during
1988 May -- 1989 June, characterised by reduction in flux of 0.6 mag in $V$ band and the
presence of absorption profiles of H$\alpha$ and He{\sc i} 6678 \AA~lines.
\citet{Roche93} identified an extended low state during 1974-77 which may be
associated with a disc loss event from the analysis of optical, infrared and X-ray
observations of X Per over a period of 25 years.
\citet{Tarasov95} reported the interesting formation of a double circumstellar
disc in X Per, inferred from the quadruple emission peak structure in He{\sc i}
6678 \AA~line.

X Per was monitored as part of a program to observe Be/X-ray binaries in
the near-IR using the 1.2m Mt. Abu telescope.
The first photometric observations indicated that X Per was  brighter by $\sim$ 0.7 mag
in $J, H, K$ compared to its listed 2MASS values providing the initial
motivation to continue observing the object.
At present, photometry and spectroscopy spanning 8 nights
spaced over a period of 3 months are reported.

Since much of this work is related to near-IR spectroscopy, it is worthwhile
to summarize the major near-IR spectroscopic studies of Be stars that
are relevant to this work. A large sample of 57 and 66 stars in the $H$ and $K$
bands respectively of spectral types O9 -- B9 and luminosity classes III, IV
and V were studied by \citet{Steele01} and \citet{Clark00}
in two studies separately devoted to the $H$ and $K$ bands respectively.
The major emphasis of these studies was on characterization of the stars based
on the lines of different species seen in their spectra. Both studies serve as
good templates for comparing or contrasting newly obtained $H$ and $K$
spectroscopic data of other Be stars.
A significant extension into understanding the $L$ band spectra of Be stars
was recently made by \citet{Granada10}. These authors used simultaneous $K$
and $L$ band spectroscopy to understand the circumstellar envelope properties
from the Brackett, Pfund and Humphrey lines of hydrogen seen in the spectra. 
From the ISO spectra, \citet{Lenorzer02} used the line flux ratio of
Hu(14)/Br$\alpha$ and Hu(14)/Pf$\gamma$ as a diagnostic tool to constrain the
geometry of the ionized circumstellar material. 
In the $J$ band, there appears to be a paucity of spectroscopic results - 
either of isolated stars or of larger samples -  although this band contains 
certain diagnostic lines of considerable physical interest as discussed in section 3.1.

\section{Observations}
The photometric and spectroscopic observations of X Per were carried out from
the 1.2m Mt. Abu telescope, operated by the Physical Research Laboratory.
The log of the photometric observations along with derived
$JHK$ magnitudes is given in Table 1.
The log of the spectroscopic observations is given in Table 2.
The near-IR $JHK$ spectra presented here were
obtained at similar dispersions of $\sim$ 9.5 \AA~/pixel in each of the
$J, H, K$ bands using the Near-Infrared Imager/Spectrometer with a
256 $\times$ 256 HgCdTe NICMOS3 array. The 1 arc second wide slit images to 2 pixels on the
detector thereby yielding a resolving power of 800--1000 in the near-IR
bands. A set of two spectra were taken with the object dithered to two positions along the slit. The
spectra were extracted using IRAF and wavelength calibration was
done using a combination of OH sky lines and telluric lines that
register with the stellar spectra. Following the standard procedure,
the object spectra were then ratioed with the spectra of a comparison
star (SAO 56762; A5V, $T_{eff}$ = 8200 K \citep{Schmidt-Kaler82}) observed at
similar airmass as the object. Prior to the ratioing process 
the hydrogen Paschen and Brackett absorption lines in the comparison stars spectrum 
are removed using a Gaussian fit using IRAF.
The ratioed spectra were then multiplied by a blackbody curve at the
effective temperature of the comparison star to yield the final spectra. Photometry
in the $JHK$ bands was done in photometric sky conditions using
the imaging mode of the NICMOS3 array. Several frames, in five
dithered positions offset typically by 20 arcsec, were obtained of
both the program object and a selected standard star (SAO 56762; A5V) in each of the $J, H, K$
filters. Near-IR $JHK$ magnitudes were then derived using IRAF tasks
and following the regular procedure followed by us for photometric
reduction (e.g. \citet{Banerjee02}).

\section{Results}

\subsection{Photometric results and general characteristics of the spectra}
The light curve of X Per taken over 6 epochs is shown in Figure 1,
indicating no significant variations over this period.
The mean near-IR brightness is high when compared with the compilation of $JHK$ magnitudes
collected over 25 years by \citet{Telting98}. For example, two of the sets of
data with lowest $J, H, K$ magnitudes (highest flux values) recorded by these authors are
5.82, 5.21 and 5.15 on 1987 August 30
and 5.44, 5.41 and 5.29 respectively between 1994 September 16 -- 20. 
This enhanced near-IR brightness could be an indication of the
accumulation of more material in the disc, resulting from episodes of stellar
mass loss events.

Regarding the spectra, the recombination emission lines of hydrogen and helium
are seen to dominate the $JHK$
spectra. The prominent lines seen are Paschen $\beta$ 1.2818 $\mu$m, Paschen
$\gamma$ 1.0938 $\mu$m
and He{\sc i} 1.0830 $\mu$m in the $J$ band (Figure 2);
Brackett 10 to 18 and He{\sc i} 1.7002 $\mu$m in the $H$
band (Figure 3) and Brackett $\gamma$ 2.1656 $\mu$m, He{\sc i} 2.058, 2.1120,
2.1132 $\mu$m in the $K$ band (Figure 4). The $H$ and $K$ band spectra of
X Per are similar to those of classical Be stars observed by
\citet{Steele01} and \citet{Clark00}.
Based on their $K$ band spectra, \citet{Clark00} classified Be stars into five
groups based on the strength and presence of the Br$\gamma$, He{\sc i} and
Mg{\sc ii} 2.138, 2.144 $\mu$m features which are seen in the $K$ band.
Group {\sc i} candidates are those stars which show Br$\gamma$ in emission along
with He{\sc i} line features which can either be in emission or in
absorption. It was also seen that all Group {\sc i}
candidates belonged to spectral class B3 or earlier. The observed presence of both Br$\gamma$
and He{\sc i} in emission in X Per would indicate that it
belongs to Group {\sc i} and its spectral type
is hence expected to be earlier than B3 --- this is consistent with its present
spectral classification of O9.5 {\sc iii}e.

An interesting aspect of the spectroscopy is the presence of the
O{\sc i} 1.1287 $\mu$m and 1.3165 $\mu$m lines in the $J$ band spectra.
The relative strengths of these lines can help discriminate whether
the Ly$\beta$ fluorescence mechanism is operational or not in the star.
The Ly$\beta$ fluorescence  mechanism was proposed by \citet{Bowen47}
wherein due to the near coincidence of wavelengths, hydrogen Ly$\beta$
photons at 1025.72 \AA~can pump the O{\sc i} ground state resonance line at
1025.77 \AA~thereby populating the O{\sc i} 3d$^3$D$^0$ level.
The subsequent downward cascade produces the 11287, 8446 and 1304 \AA~lines in
emission thereby enhancing the strengths of these lines.
It is expected that $W$(1.3165)/$W$(1.1287) $\ge$  1 if continuum fluorescence is the significant
excitation mechanism for these lines ( \citet{Strittmatter77},
\citet{Grandi75}; $W$ is the equivalent width). On the other hand, if excitation by the Ly$\beta$
fluorescence process is significant, the 1.1287 $\mu$m line should
become stronger of the two lines. Since we measure a  mean
value of $W$(1.3165)/$W$(1.1287) = 0.43 for
for all epochs of our observations, it is implied that the
Ly$\beta$ fluorescence  process
is operative and has a significant role in the excitation of the O{\sc i} lines.

The equivalent widths of the prominent lines, measured in \AA~are given in Table 3 -- the typical
error in the measurement of the equivalent width values is 10 \%.

\begin{figure}
\includegraphics[width=84mm]{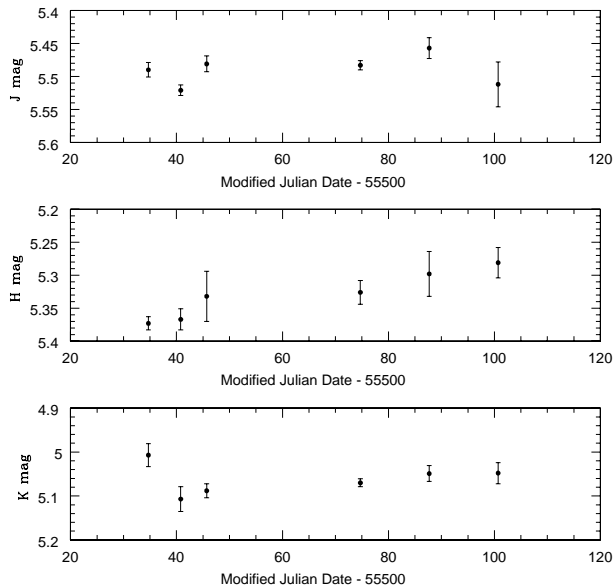}
\caption{Near-IR light-curve of X Per during the period 2010 December to 2011 February.}
\end{figure}

\begin{table}
 \centering
\caption{Journal of the photometric observations.
The errors on the $JHK$ magnitudes are shown in brackets. }
\begin{tabular}{@{}lcccc@{}}
\hline
 Date & MJD & \multicolumn{3}{c}{Photometry}\\
      &     & \multicolumn{3}{c}{Magnitudes}\\
      &     & $J$ & $H$ & $K$ \\
\hline
2010 &  &  & & \\
Dec. 04 & 55534.7 & 5.49 (0.01) & 5.37 (0.01) & 5.01 (0.02) \\
Dec. 10 & 55540.8 & 5.52 (0.01) & 5.37 (0.02) & 5.11 (0.03) \\
Dec. 15 & 55545.7 & 5.48 (0.01) & 5.33 (0.04) & 5.09 (0.02) \\
2011 &  &  & &  \\
Jan. 13 & 55574.7 & 5.48 (0.01) & 5.33 (0.02) & 5.07 (0.01)  \\
Jan. 26 & 55587.7 & 5.46 (0.02) & 5.30 (0.03) & 5.05 (0.02) \\
Feb. 08 & 55600.7 & 5.51 (0.03) & 5.28 (0.02) & 5.05 (0.02) \\
\hline
\end{tabular}
\end{table}

\begin{table}
 \centering
\caption{Journal of the spectroscopic observations.}
\begin{tabular}{@{}lcccccc@{}}
\hline
 Date & \multicolumn{3}{c}{Spectroscopy}& \multicolumn{2}{c}{Airmass}\\
      & \multicolumn{3}{c}{Exp.time (s)}& X Per & SAO 56762\\
      &  $J$ & $H$ & $K$ & ($J$, $H$, $K$) & ($J$, $H$, $K$) \\
\hline
2010 &  &  &  &  & \\
Dec. 05 & 40 & 40 & 50 & (1.01,1.02,1.05) & (1.13,1.17,1.19)\\
Dec. 10 & 50 & 40 & 50 & (1.14,1.16,1.19) & (1.12,1.18,1.20)\\
Dec. 15 & 90 & 60 & 60 & (1.08,1.10,1.22) & (1.08,1.12,1.19)\\
2011 &  &  &  &  & \\
Jan. 13 & 60 & 60 & 60 & (1.09,1.10,1.13) & (1.09,1.11,1.16)\\
Jan. 26 & 60 & 40 & 40 & (1.14,1.16,1.19) & (1.14,1.18,1.20)\\
Jan. 28 & 40 & 40 & 40 & (1.33,1.28,1.25) & (1.36,1.30,1.25)\\
Feb. 08 & 60 & 60 & 120 & (1.06,1.05,1.04) & (1.06,1.05,1.03)\\
\hline
\end{tabular}
\end{table}

\begin{figure}
\includegraphics[width=84mm]{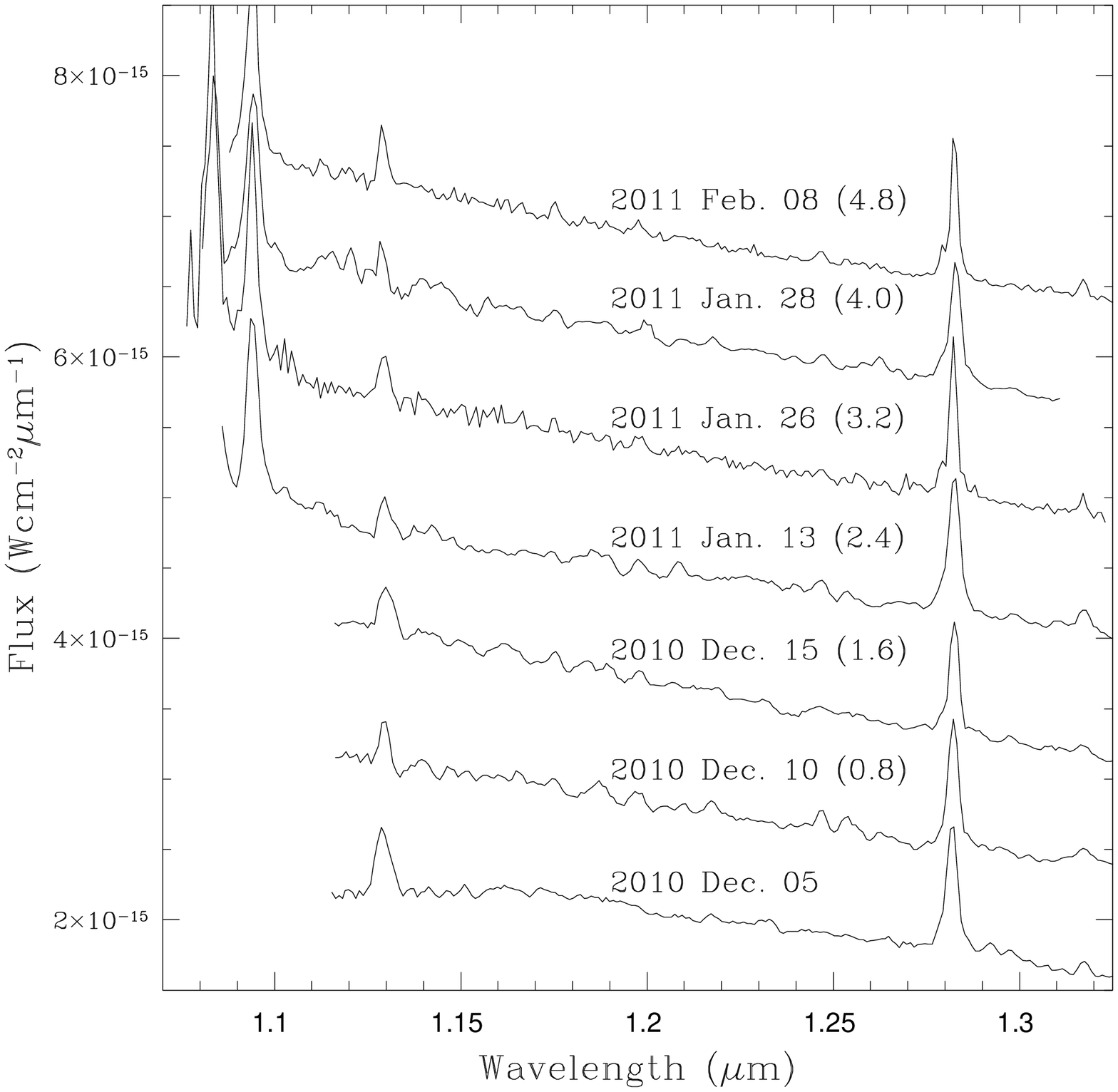}
\caption{Flux calibrated $J$ band spectra of X Per are displayed at different
  epochs with an offset between adjacent spectra for clarity. The amount of offset in
  units of 10$^{-15}$ Wcm$^{-2}$$\mu$m$^{-1}$ is shown in brackets after the
  date of observation.}
\end{figure}

\begin{figure}
\includegraphics[width=84mm]{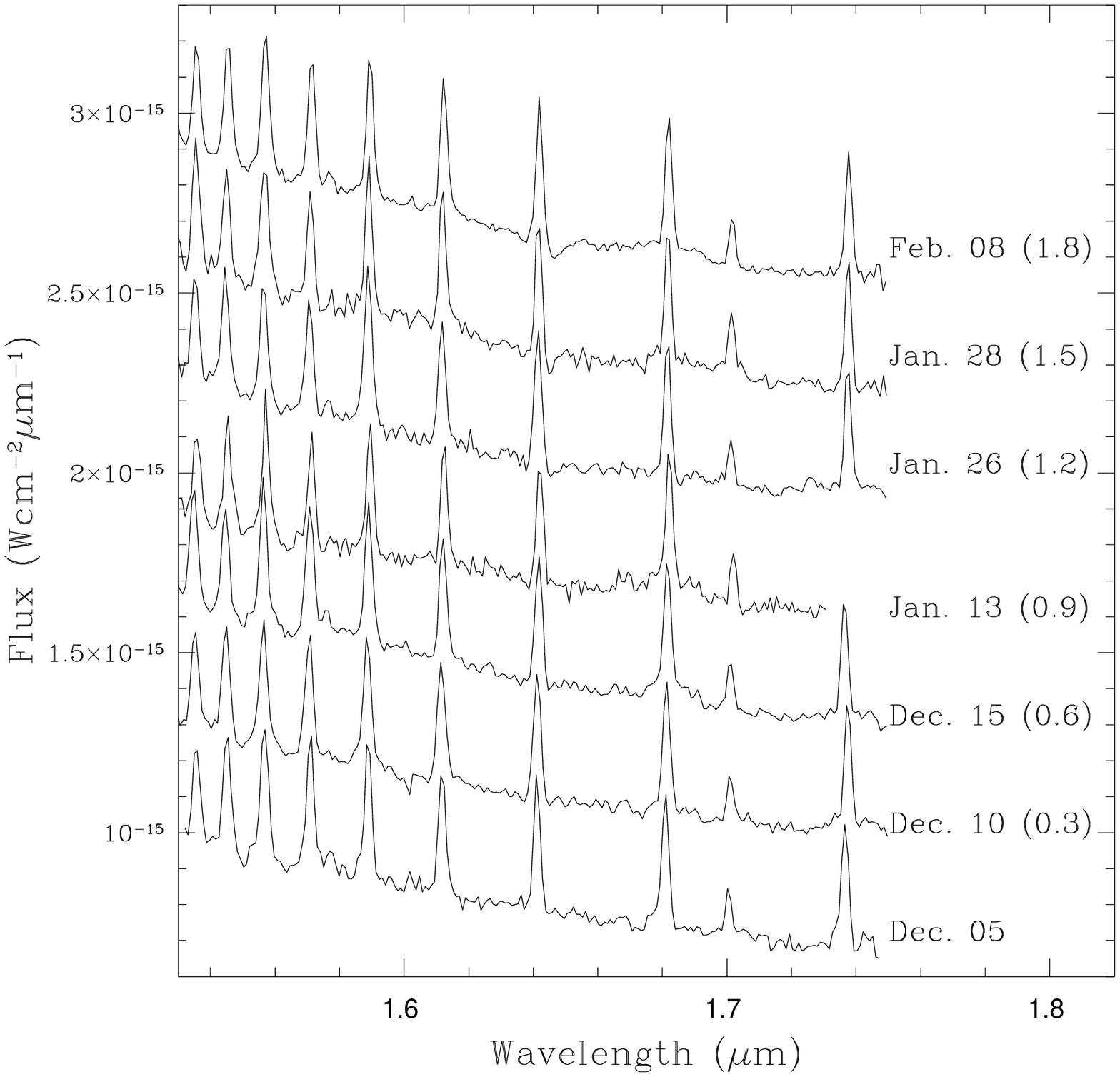}
\caption{Flux calibrated $H$ band spectra of X Per are displayed at different
  epochs with an offset between adjacent spectra for clarity. The amount of offset in
  units of 10$^{-15}$ Wcm$^{-2}$$\mu$m$^{-1}$ is shown in brackets after the
  date of observation.}
\end{figure}

\begin{figure}
\includegraphics[width=84mm]{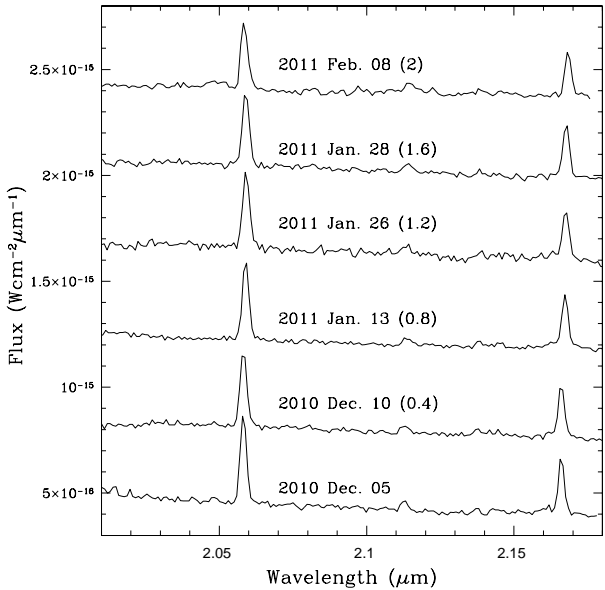}
\caption{Flux calibrated $K$ band spectra of X Per are displayed at different
  epochs with an offset between adjacent spectra for clarity. The amount of offset in
  units of 10$^{-15}$ Wcm$^{-2}$$\mu$m$^{-1}$ is shown in brackets after the
  date of observation.}
\end{figure}

\subsection{Analysis of the continuum}
We construct and analyse the spectral energy distribution of the object in this section
based on the near-IR magnitudes.
Since the $JHK$ photometric estimates during the period of observation do not
change much (Table 1), it is adequate to model the SED by considering
the  data for one representative epoch -
we have taken the 2010 December 10 data for this purpose. Figure 5 shows the SED
where the $JHK$ spectra have been dereddened using the corresponding $JHK$
photometric values and $E(B-V)$ = 0.39 \citep{Fabregat92} using the task DEREDDEN in IRAF.
We have shown a  blackbody curve at T = 31400 K corresponding to the
effective temperature determined for X Per \citep{Fabregat92}.
In spite of a thorough search of available databases, we are unable to locate
any $V$ band measurement contemporaneous with our near-IR observations.
We have thus used the long-term compilation of photometric
magnitudes of \citet{Telting98} and assume that the colors of the star
(for e.g. $(V-J)$, $(V-H)$ etc.) should remain fairly
the same at similar brightness levels. That is, similar $JHK$ magnitudes in Telting's
and the present study should be accompanied by similar $V$ magnitudes.
As mentioned earlier, our $JHK$ values are similar to that
obtained on 1994 September 16 -- 20 when X Per was in a high brightness state.
Hence we have taken the $V$ magnitude as 6.24, corresponding to 1994 September
25, as given in \citet{Telting98}. The blackbody curve in Figure 5 has hence
been anchored to this $V$ band magnitude  which is also deredenned using
$E(B-V)$ = 0.39. From Figure 5 it is evident that a blackbody curve poorly fits the SED of X Per and
an infrared excess is seen which we attribute to free-free (f-f) emission from the disc.

The observed free-free excess can be modeled to obtain an average value of the electron density
in the disc. Given a distance $D$ to the object, a volume $v$ for the emitting
ionized gas of the disc, the observed flux $F$
(in units of W cm$^{-2}$ $\mu$m$^{-1}$ ) due to f-f contribution will be given by

\begin{equation}
F = j_{\lambda ff} \times v / 4 \pi D^2
\end{equation}

where the free-free volume emission coefficient, $j_{\lambda ff}$
(in units of W cm$^{-3}$ $\mu$m$^{-1}$) can be calculated from

\begin{equation}
j_{\lambda ff} = 2.05\times10^{-30} \lambda^{-2} z^2 g T_s^{-1/2} n_e n_i exp(-c2 / \lambda T_s)
\end{equation}

In the above $\lambda$ is the wavelength of emission in $\mu$m, $z$ is the
charge, $g$ is the Gaunt factor, $T_s$ is the disc temperature, $n_e$
and $n_i$ are the electron and ion densities respectively and 
$c2$ = 1.438 cm K \citep{Banerjee01}.
In the case of a circumstellar disc, the volume of the emission
region ($v$) is reasonably estimated as $\pi$$R_s$$^2$$H_s$, where $R_s$ and $H_s$ are the disc
radius and thickness respectively. In the case of Be stars, the disc thickness
can be approximated to be one-fifth of stellar radius \citep{Gehrz74}. We assume
$g$ and $z$ to be unity, $n_e$ = $n_i$ for a pure hydrogen shell, and adopt a
distance to the object of 1300 pc from \citet{Fabregat92}.
The calculated values of free-free electron flux as a function of wavelength is
shown as dotted line in Figure 5 for a temperature of 10,000 K and
$n_e$ = 4$\times$10$^{11}$ cm$^{-3}$. The free-free contribution,
computed for a choice of these parameters, when added to the
blackbody curve is found to reproduce
the observed SED much better than a blackbody alone. This value of electron
density is comparable with that expected from observational modeling 
(e.g. \citet{Silaj10}; \citet{Carciofi08}; \citet{Gies07}). 
A realistic model should take into account the optical depth effects while
calculating the continuum emission.

\begin{figure}
\includegraphics[width=84mm]{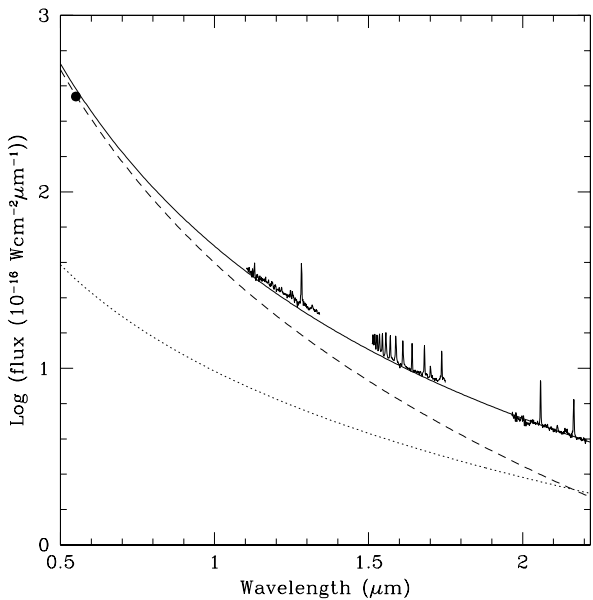}
\caption{The spectral energy distribution of X Per for the epoch 2010 December 10
  is shown in the figure. The flux calibrated $JHK$ spectra are dereddened
  using $E(B-V)$ = 0.39. The blackbody corresponding to the central stars temperature of 
  T = 31400 K is shown by a 
  dashed line, the free-free contribution from the disc  by a  dotted line
and their co-added sum by a  solid line. Also shown is the $V$ band flux
corresponding to high brightness state (filled circle). For further details see section
3.2. }
\end{figure}

\begin{figure}
\includegraphics[width=84mm]{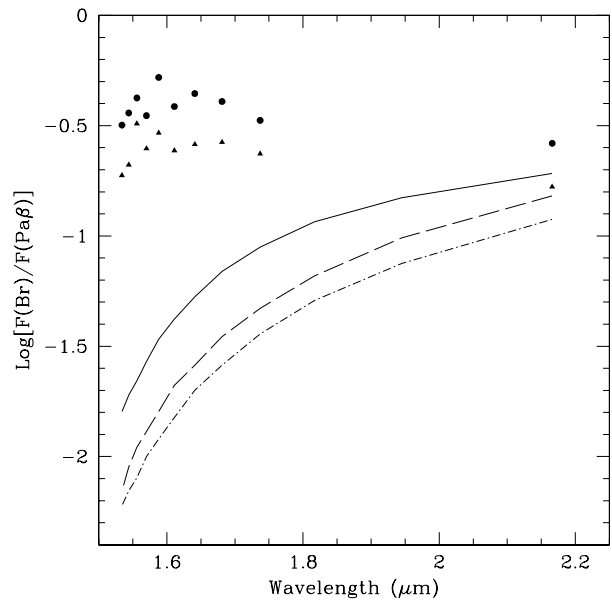}
\caption{Recombination case B analysis of the Brackett lines of hydrogen are shown for X Per
where the Br line fluxes are all normalized with respect to the line flux of Pa $\beta$.
The lines shown are Br10 -- 18 (1.7367 -- 1.5346 $\mu$m) and Br$\gamma$.
A large deviation from case B predictions is seen especially in the behavior of the higher Br lines.
Filled triangles indicate the observations taken on 2010 December 10 while the filled
circles correspond to that on 2011 January 26.
Solid lines represent case B values for $T_e$ = 10$^4$ K,
$n_e$ = 10$^{10}$ (shown as continuous line), 10$^{12}$ (shown as dashed line)
and 10$^{14}$ cm$^{-3}$ (shown as dot-dashed line).}
\end{figure}

\begin{table*}
 \centering
\caption{Measured equivalent widths of the emission lines in \AA.}
\begin{tabular}{@{}lrrrrrrrrrrrrrrr@{}}
\hline
Date & Pa$\beta$ & Pa$\gamma$ & He{\sc i} & Br10 & He{\sc i} & Br11 & Br12 &
Br13 & Br14 & Br15 & Br16 & Br17 & Br18 & Br$\gamma$ & He{\sc i} \\
     & & & (1.08) & & (1.70) & & & & & & & & & & (2.05) \\
     & & & ($\mu$m) & & ($\mu$m) & & & & & & & & & & ($\mu$m)\\
\hline
2010 & & &  & &  & & & & & & & & & & \\
Dec. 05 & 14.3 & -- & -- & 15.4 & 3.6 & 15.2 & 11.1 & 12.0 & 11.2 & 7.5
& 9.8 & 9.6 & 4.3 & 15.1 & 23.7 \\
Dec. 10 & 21.0 & -- & -- & 13.1 & 4.1 & 14.6 & 13.5 & 11.6 & 12.9
& 10.8 & 13.1 & 8.0 & 7.7 & 20.8 & 23.8 \\
Dec. 15 & 14.3 & -- & -- & 11.0 & 4.7 & 13.0 & 12.7 & 10.7 & 11.7 & 11.1
& 11.8 & 8.4 & 8.2 & -- & -- \\
2011 & & &  & &  & & & & & & & & & & \\
Jan. 13 & 22.6 & 18.4 & -- & -- & 4.1 & 14.9 & 11.2 & 9.9 & 11.4 & 11.3 & 13.5 & 10.3
& 7.3 & 19.7 & 24.0 \\
Jan. 26 & 16.8 & 15.9 & 28.9 & 11.4 & 4.0 & 13.5 & 14.1 & 11.4 &
14.2 & 9.6 & 10.0 & 9.4 & 7.8 & 18.1 & 22.1 \\
Jan. 28 & 12.8 & 17.0 & 19.0 & 13.1 & 4.1 & 10.8 & 12.1 & 10.2 &
10.8 & 7.4 & 12.2 & 7.4 & 9.8 & 16.2 & 20.6 \\
Feb. 08 & 17.4 & 20.4 & -- & 11.8 & 4.1 & 11.3 & 12.7 & 10.4 & 11.6 & 9.6
& 11.7 & 9.1 & 7.2 & 14.3 & 22.8 \\
\hline
\end{tabular}
\end{table*}

\subsection{ Recombination analysis of the hydrogen lines}
Of the H{\sc i} lines, only two of the Paschen series lines could be covered in the
spectra presented here viz. Pa$\beta$ at 1.2818 $\mu$m and Pa$\gamma$ at
1.0938 $\mu$m. Whenever recorded, Pa$\gamma$ is found to be stronger than
Pa$\beta$  contrary to what is expected (see Figure 2). This indicates that
these Paschen lines are optically thick
since the expected ratio in recombination case B conditions is
$I$(Pa$\beta$)/$I$(Pa$\gamma$) $\sim$ 1.57 -- 2.01 for typical densities and
temperatures prevailing in Be star discs (i.e. $T_e$ = 10$^4$ K, $n_e$ in the range
10$^{10}$ to 10$^{14}$cm$^{-3}$; here $I$ is the line intensity in units
of erg cm$^{-2}$ s$^{-1}$ whose values are taken from \citet{Storey95}).
In essence, Pa$\beta$ is always expected to be stronger
than Pa$\gamma$ under optically thin case B conditions. Optical depth effects are more
clearly seen in the Brackett series lines viz. Br$\gamma$ in the $K$ band
and Br10 to 18 in the $H$ band. In Figure 6, we present plots of
the observed strength of  Br lines versus their predicted
intensities under recombination case B conditions. The line fluxes used in
this figure were derived from the spectra which were flux calibrated by the
broad-band $JHK$ magnitudes of Table 1 and dereddened using a value of $A_V$
= 1.19 \citep{Valencic08}.

We have presented the data only
for 2010 December 10 and 2011 January 26 in Figure 6
since there is considerable cluttering and loss of clarity if
the observed data of all 5 days are presented. But we found that a similar trend for the line
strengths, as presented in Figure 6, is seen too for the data of the other
days. The case B line intensities
are from \citet{Storey95} for a  temperature $T_e$ = 10$^4$ K and for
three representative values of the electron density
$n_e$ = 10$^{10}$, 10$^{12}$ and 10$^{14}$ cm$^{-3}$, respectively. As can be
seen from Figure 6, the observed line intensities deviate significantly from the
optically thin case B values. This indicates that Brackett lines are optically
thick during all the epochs of observation of X Per.
This observed behaviour of the Br line strengths is consistent with that seen in Be stars in general.
\citet{Steele01}, from their $H$ band spectroscopy of 57 Be stars, showed that the strengths
of Br 11 to Br 18, relative to each other, do not in general fit case B theory
particularly well. Being a paper devoted to the $H$ band, they did not include
Br$\gamma$ in their analysis, but its inclusion here in Figure 6 brings out
the deviation from case B values even more clearly.
From recombination theory it is qualitatively expected, under optically thin conditions, 
 that when  strengths of lines of the same series are compared, a lower line of the series should  
 be stronger
 than a higher line. For example, it is  expected
that Br$\gamma$ (corresponding to a transition between levels 7--4) is expected to be stronger than
any higher line of the series like Br10 or Br11 (transitions between 10---4
and 11--4 respectively). 
But the reverse is actually being observed here.

It is interesting  to note that lines of even the higher Pfund (Pf) and
Humphrey (Hu) series could in general be optically thick in Be stars.
For e.g., of the 8 Be stars for which the  $L$ band spectra have been presented
by \citet{Granada10}, the spectra of EW Lac and BK Cam cover several of
the Pf and Hu lines viz. Pf 8, 9, 10 and 17 -- 27 \& Hu 14 -- 25.
Analysis of their strengths, and comparison with case B predictions, shows a
strong departure from the optically
thin case (Figures 4 and 5 in \citet{Granada10}).

We analyzed the case B model values to check whether opacity effects are expected to affect
the strengths of the Br line shown in Figure 6 under the density conditions prevailing in Be discs.
It is verified  from  \citet{Hummer87} and \citet{Storey95} that
line center optical depth values can be  significant when the densities
become large  as in Be star discs.
The above studies tabulate the value of the opacity factor $\Omega$$_{n,n'}$ for different H{\sc i}
lines for transitions between levels $(n,n')$ at different temperatures and
densities (equation 29 of \citet{Hummer87}). From this value of $\Omega$$_{n,n'}$, 
the optical depth at line-center $\tau$$_{n,n'}$ is to be calculated using
$\tau$ = $n_e$$n_i$$\Omega$$L$, where $L$ is the path length in cm.
As a representative  example we consider a Br line photon,
originating in the disc, trying to escape across the thickness of the disc. Then $L$ may reasonably
be approximated as the thickness of the circumstellar disc which is typically assumed to be
1/5 times the stellar radius. For X Per, the stellar radius is estimated
as 13$\times$$R_\odot$ \citep{Fabregat92} for which $L$ comes out
to be 1.82$\times$10$^{11}$ cm. The corresponding value of the
optical depth $\tau$, from \citet{Storey95}, is then found to be  $\sim$ 13500
for Br$\gamma$ and this then decreases monotonically down the series to a
value of $\tau$ $\sim$ 300 for Br 18 (these $\tau$ values
are computed at representative values of $T_e$ = 10$^4$ K and $n_e$ =
10$^{13}$ cm$^{-3}$). At a lower density  of $n_e$ = 10$^{11}$ cm$^{-3}$,
the $\tau$ values decrease by approximately a factor
of $\sim$ 10000 (since $\tau$ $\propto$ $n_en_i$) to $\tau$ $\sim$ 1
so lines like Br$\gamma$ still remain significantly thick.
Thus recombination theory does predict that the Br and Pa lines should in
general be optically thick. Going beyond the Pa and Br lines,
it may also be easily verified from the \citet{Storey95} data
that the optical depth in the Humphrey and
Pfund lines are also large - of similar magnitude as the $\tau$ values for the
Br lines - at similar densities considered here.
This is likely to explain the observed deviations of the strength of
these lines from case B predictions as seen in the data of \citet{Granada10}.

Unrelated to Be stars but worth mentioning in this connection, is the analogous behaviour of
the Pa and Br line strengths arising from the ionized ejecta of novae.
Here too, similar optical depth effects are sometimes seen as for example
in the spectra of nova Oph 1998 (V2487 Oph) and nova Sgr 2001 (V4643 Sgr) studied by
\citet{Lynch00} and \citet{Ashok06}, respectively. The large
optical depths seen in the  Br and Pa lines in such cases are well explained by a model
developed by \citet{Lynch00} who show  that the relatively larger intensities of the
higher members of the Paschen and Brackett series arise because of emission
from high-density or optically thick emission line gas. In effect, it is only
at high densities (around 10$^{10}$ -- 10$^{12}$)
which occur in novae ejecta just after outburst before expansion dilutes the
ejecta, that optical depth effects become pronounced.

\section{Discussion}
In Be stars the situation regarding optical depth effects could be
complex since the optical depth in a line will also depend on
the region from where the line emanates.
Interferometric results indicate that different lines originate from different regions
in the circumstellar disc, i.e., regions of different electron density and
hence different $\tau$ values. The presence of density enhancements due to
spiral waves in the disk could complicate matters further
(\citet{Wisniewski07}, \citet{Hesselbach09}\footnote{http://utmost.cl.utoledo.edu}). 
Disc size estimates that are now available show considerable variation in sizes from
measurements around the H$\alpha$ emission line
(e.g. \citet{Quirrenbach97}, \citet{Stee05}; several papers by Tycner and
collaborators); in the infrared 10 $\mu$m $N$ band continuum \citep{Chesneau05};
in the near-IR $K$ band (e.g. \citet{Gies07}); in the $H$ band continuum and
also in $K$ band Br$\gamma$ and
He{\sc i} 2.058 $\mu$m emission lines \citep{MillanGabet10}.
A summary of the interferometric results available upto fairly recent times
may be found in \citet{Gies07} and Monnier (2003; and references therein).
Variations in disc sizes, though not measured directly, are also suggested by
the modeling of \citet{Jaschek93} for the O{\sc i} and Paschen lines.

\section*{Acknowledgments}
We thank the referee, Prof. Douglas R. Gies, for his suggestions and 
comments which helped in improving the manuscript. 
The research work at Physical Research Laboratory is funded by the Department
of Space, Government of India.

\end{document}